\documentclass{article}
\usepackage{spconf,amsmath,graphicx,hyperref}
\usepackage{comment, url, algorithm, algpseudocode, makecell, multirow, amssymb}

\title{SAGA-SR: Semantically and Acoustically Guided Audio Super-Resolution}
\name{Jaekwon Im \qquad Juhan Nam\thanks{This work was supported by the National Research Foundation of Korea (NRF) grant funded by the Korea government (MSIT) (No. RS-2023-00222383).}}
\address{Graduate School of Culture Technology, KAIST, Republic of Korea}
\begin{document}
%\ninept
%
\maketitle
\begin{abstract}
Versatile audio super-resolution (SR) aims to predict high-frequency components from low-resolution audio across diverse domains such as speech, music, and sound effects. Existing diffusion-based SR methods often fail to produce semantically aligned outputs and struggle with consistent high-frequency reconstruction. In this paper, we propose SAGA-SR, a versatile audio SR model that combines semantic and acoustic guidance. Based on a DiT backbone trained with a flow matching objective, SAGA-SR is conditioned on text and spectral roll-off embeddings. Due to the effective guidance provided by its conditioning, SAGA-SR robustly upsamples audio from arbitrary input sampling rates between 4 kHz and 32 kHz to 44.1 kHz. Both objective and subjective evaluations show that SAGA-SR achieves state-of-the-art performance across all test cases. Sound examples and code for the proposed model are available online\footnote{\url{http://jakeoneijk.github.io/saga-sr-project}}.
\end{abstract}
\section{Introduction}
\label{sec:intro}
%%%%% Audio super-resolution이란 %%%%%
Audio super-resolution (SR) aims to reconstruct a high-resolution audio signal from its corresponding low-resolution audio signal. To enhance listening experiences, it can be applied to diverse audio types, including historical recordings, low-bandwidth telephone audio, and audio generated by deep learning models \cite{audiosr}. Previous audio SR methods \cite{earlysr, musicsr} based on deep neural networks have achieved promising performance in constrained settings, including fixed upsampling ratios and restricted domains such as speech or music. However, these constraints limit their applicability in diverse and complex real-world scenarios.

To address this issue, several recent works \cite{audiosr,flashsr} have focused on versatile audio super-resolution, which is the task of upsampling general-domain audio, including music, speech, and sound effects, from varying input sampling rates to full bandwidth, such as 44.1 kHz or 48 kHz. AudioSR \cite{audiosr} employs a latent diffusion model (LDM) \cite{stablediffusion} with a Transformer-UNet backbone \cite{audioldm2} to capture the complex distributions of general audio signals. %The LDM predicts the VAE \cite{vae} latent, which is transformed into a mel-spectrogram by the VAE decoder and subsequently converted into waveforms by HiFi-GAN \cite{hifigan}.
FlashSR \cite{flashsr} employs diffusion distillation \cite{flashdiffusion} to train the Student LDM using AudioSR as the Teacher LDM and proposes the SR Vocoder to further enhance AudioSR’s performance.

Although previous methods have achieved notable success, there is still room for improvement. There are two key challenges hindering the performance of versatile audio SR models. First, to generate natural high-resolution audio, an audio SR model needs to capture semantic information from low-resolution input and effectively incorporate it into the reconstruction process. Previous methods often fail to predict semantically-aligned high-frequency components, resulting in unnatural artifacts, such as excessive sibilance \cite{flashsr}. Second, unlike speech SR, versatile audio SR handles a much broader range of audio domains, which exhibit high diversity in high-frequency energy distributions. This results in difficulties for models in consistently reconstructing high-frequency content, especially when the input has a low cutoff-frequency (e.g., 4 kHz).

In this paper, we present SAGA-SR, a versatile audio super-resolution model that leverages semantic and acoustic conditions. SAGA-SR is based on a DiT \cite{dit} backbone trained with a flow matching objective \cite{flowmatching}, incorporating two key conditions. First, inspired by recent works \cite{textimagesr1, textvideosr} in the computer vision domain, we utilize text embeddings for semantic guidance. Specifically, we employ an audio-language model to generate text captions from audio, enabling more efficient training and inference. Second, we introduce spectral roll-off embeddings, which provide relative high-frequency energy information for both the input and target audio. Guided by both semantic and acoustic conditions, SAGA-SR can robustly upsample music, speech, and sound effects from any sampling rate between 4 kHz and 32 kHz to 44.1 kHz, and achieves state-of-the-art performance on both objective and subjective evaluations.

\section{Method}
\label{sec:methodology}

\begin{figure}[!t]
\centering
\begin{minipage}{0.95\columnwidth}
  \centering
\centerline{\includegraphics[width=\columnwidth]{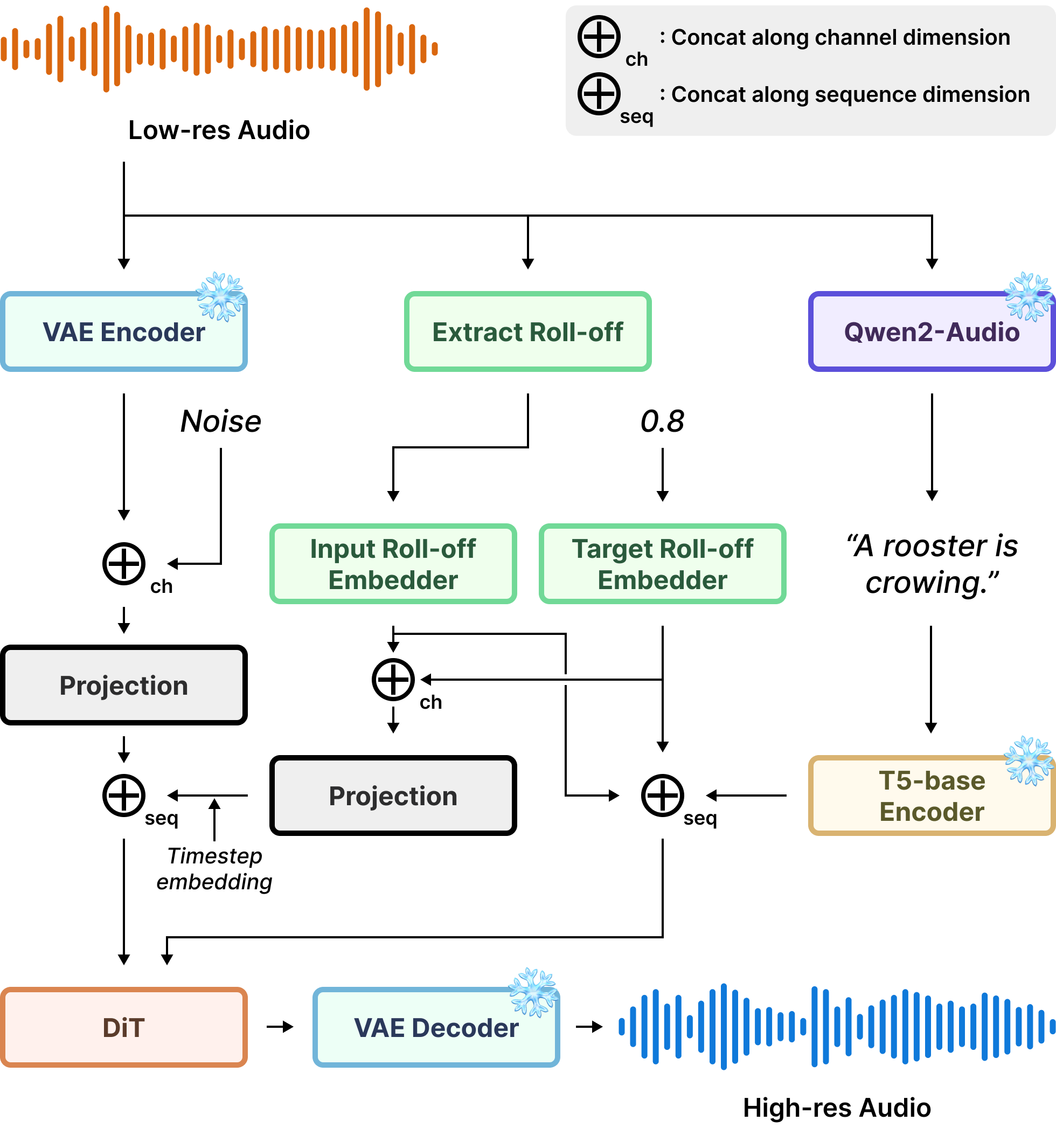}}
\end{minipage}
\caption{Overview of SAGA-SR}
\label{fig:model}
\end{figure}

Figure \ref{fig:model} shows the overall architecture of SAGA-SR. Let $x_h$, $x_l \in \mathbb{R}^{2 \times L}$ represent the high-resolution and low-resolution audio, respectively, where $L$ is the number of samples. Each audio sample is compressed into latent representations $z_h\in \mathbb{R}^{64 \times L/2048}$ and $z_l \in \mathbb{R}^{64 \times L/2048}$ by the pre-trained VAE encoder from \cite{stableaudioopen}. The text $c$ describes the audio content independent of its resolution. Roll-off frequencies $f_h$, $f_l \in \mathbb{R}$ are extracted from $x_h$ and $x_l$, respectively. DiT estimates $z_h$ from $z_l$, $c$, $f_h$, and $f_l$. The predicted latent is then converted into an audio signal by the pre-trained VAE decoder, followed by low-frequency replacement post-processing \cite{audiosr} to ensure consistency in low-frequency information. In Section \ref{subsec:dit_train}, we present the training and inference procedures of DiT. In Section \ref{subsec:conditioning}, we describe how each condition is processed to be incorporated into DiT.

\subsection{DiT Model}
\label{subsec:dit_train}
Unlike previous works \cite{audiosr, flashsr} that employ a Transformer-UNet architecture, we use DiT \cite{dit}, which is widely adopted in image and audio generation models. We adapt the DiT architecture proposed in \cite{stableaudioopen} and train it using the conditional flow matching objective \cite{flowmatching}. The flow matching objective regresses onto a target vector field that generates a probability path, transforming a simple distribution into an approximation of the data distribution. We use a linear interpolation path between the noise and the data as follows:
\begin{equation}
    z_t=(1-t) \cdot z_0 + t\cdot z_1,
\end{equation}
where $z_1=z_h$, $z_0 \sim \mathcal{N}(0,1)$, and $t \in [0,1]$. The corresponding velocity at $z_t$ is given by
\begin{equation}
v_t = \frac{dz_t}{dt}= z_1 - z_0.
\end{equation}
The training objective for DiT is defined as
\begin{equation}
    \mathbb{E}_{t,z_0,z_1,z_l,c}  \lVert u(z_t,z_l,c, f_h, f_l, t;\theta) - v_t\rVert^{2},
\end{equation}
where $\theta$ denotes the model parameters. To condition DiT on $z_l$, we concatenate $z_l$ with $z_t$ along the channel dimension. A dropout rate of 10\% is applied to $z_l$, enabling classifier-free guidance \cite{cfg}.

At inference, we sample $z_0 \sim \mathcal{N}(0,1)$ and use an ODE solver to generate $\hat{z}_h$. In practice, we adopt the Euler sampler with a linear-quadratic t-schedule \cite{moviegen} and 100 inference steps. To control the influence of $z_l$ and $c$, we adopt the classifier-free guidance method introduced in \cite{multicgf}.
\begin{equation}
\begin{split}
    &u_{\text{CFG}}(z_t, z_l, c, f_h, f_l, t; \theta) = u(z_t, \varnothing, \varnothing, f_h, f_l, t; \theta)\\
    &+ s_a(u(z_t, z_l, \varnothing, f_h, f_l, t; \theta) - u(z_t, \varnothing, \varnothing, f_h, f_l, t; \theta)) \\
    & + s_t(u(z_t, z_l, c, f_h, f_l, t; \theta) - u(z_t, z_l, \varnothing, f_h, f_l, t; \theta)),
\end{split}
\end{equation}
where $s_a$ and $s_t$ are guidance scales, and $\varnothing$ denotes null conditioning. We empirically set $s_a=1.4$ and $s_t=1.2$.

\subsection{Conditioning}
\label{subsec:conditioning}
\textbf{Text embedding.}
Audio super-resolution models are typically trained with high-resolution audio-only data, since low-resolution audio can be simulated from it. Compared to audio-only data, audio-text data is expensive to curate at scale. Furthermore, relying on user-provided text during inference can limit practical applicability and potentially degrade generation quality. To address these issues, we employ Qwen2-Audio \cite{qwen2} to generate text captions from audio. During training, captions generated from high-resolution audio are used for efficiency, while at inference, they are derived from low-resolution audio. Text embeddings are extracted from the generated captions using a pretrained T5-base encoder \cite{t5} and provided to the DiT through cross-attention. A dropout rate of 10\% is applied to the text embeddings. \vspace{2mm} 

\noindent \textbf{Spectral roll-off embedding.}
We compute the roll-off frequency from the STFT spectrogram using an open-source method\footnote{\url{https://librosa.org/doc/0.11.0/generated/librosa.feature.spectral_rolloff.html}}. Instead of computing it frame-wise as in the original implementation, we sum over the time axis of the magnitude spectrogram to obtain a single roll-off frequency value for each audio sample, which is then normalized to $[0,1)$ using the min–max normalization. The normalized roll-off frequency value is projected into learnable Fourier embeddings. We extract the spectral roll-off embeddings from both low-resolution and high-resolution audio. 

The spectral roll-off embeddings are conditioned into the DiT through two mechanisms. First, they are concatenated with the text embeddings along the sequence dimension before cross-attention. Second, the input and target roll-off embeddings are concatenated along the channel dimension, projected by linear layers, summed with the timestep sinusoidal embeddings \cite{ddpm}, and then prepended to the input of DiT. The input roll-off embeddings provide the DiT with information about the cutoff frequency of the input audio, improving its ability to handle varying input sampling rates, while the target roll-off embeddings guide the amount of high-frequency energy to generate. During inference, the target normalized roll-off frequency serves as conditioning, enabling the user to control the high-frequency energy in the generated audio. Because it is represented as a single scalar in $[0,1)$, it is straightforward to manipulate.

\section{Experiments}
\label{sec:experiments}
\subsection{Training Dataset and Preprocessing}
Our training dataset configuration and data simulation method are consistent with previous works \cite{audiosr,flashsr}. We train on the FreeSound \cite{freesound}\footnote{\url{https://labs.freesound.org/}}, MedleyDB \cite{medleydb}, MUSDB18-HQ \cite{musdb18}, MoisesDB \cite{moisedb}, and OpenSLR\footnote{\url{https://openslr.org/}} speech dataset \cite{speechdata}, with a total audio duration of around 3,800 hours. All audio was resampled to 44.1 kHz and randomly segmented into 5.94-second clips for training. To simulate low-high resolution audio pairs, we apply low-pass filtering to the high-resolution audio. The cutoff frequency is uniformly sampled between 2 kHz and 16 kHz. The low-pass filter type is randomly selected from Chebyshev, Butterworth, Bessel, and Elliptic, with the filter order chosen between 2 and 10.
\subsection{Implementation Details}
The DiT was trained for 26,000 steps using the AdamW optimizer with $\beta_1 = 0.9$ and $\beta_2 = 0.999$. We use a batch size of 256 and a learning rate of $1.0 \times 10^{-5}$. An InverseLR scheduler \cite{stableaudioopen} is applied with an inverse gamma of $10^{6}$, a power of 0.5, and a warmup factor of 0.99. To compute the roll-off frequency, we extract the STFT spectrogram using a Hann window of 2048 and a hop size of 512. The roll-off percentage is set to 0.985.
\subsection{Evaluation}
\noindent \textbf{Comparison.}
We compare SAGA-SR against state-of-the-art models, AudioSR \cite{audiosr} and FlashSR \cite{flashsr}. The official implementations and checkpoints are used for all comparison models. In addition, we conducted an ablation study to evaluate the effectiveness of the text embedding and the spectral roll-off embedding. We train two SAGA-SR variants, one without the text embedding and another without the spectral roll-off embedding. Both models are trained under the same settings as SAGA-SR.

\noindent \textbf{Dataset.}
For both objective and subjective evaluations, we adopt the VCTK test set (speech) \cite{nvsr}, FMA-small (music) \cite{fmasmall}, and ESC50 fold-5 (sound effects) \cite{esc50}. We selected 400 samples from each dataset.

\noindent \textbf{Evaluation Metrics.}
For objective evaluation, each sound category is evaluated at cutoff frequencies of 4 kHz and 8 kHz. Log-Spectral Distance (LSD) is used as the evaluation metric, following previous studies \cite{audiosr, flashsr, nvsr}. Although the LSD metric is widely used in audio super-resolution tasks, it has been found that it does not always align with perceptual quality \cite{audiosr, flashsr}. To complement LSD, we also adopt the Fréchet Distance (FD) based on OpenL3 \cite{openl3} for evaluating the music and sound effects categories. FD compares the statistics of embeddings from generated audio with those from ground-truth audio.

For subjective evaluation, a listening test was conducted with 25 participants. For all sound categories, the cutoff frequency was set to 4 kHz. During the test, participants were provided with the low-resolution input audio as a low anchor and asked to rate the perceptual quality of the outputs from each model on a scale from 1 to 5. We evaluated AudioSR, FlashSR, and the proposed SAGA-SR.

\section{Results}
\label{sec:results}

\subsection{Objective Evaluation}
\begin{table*}[t]
\caption{Objective evaluation results. Bold numbers indicate the best performance, while underlined numbers denote the second-best performance across all models. Note that low-frequency replacement post-processing was applied to the audio reconstructed by the VAE.}
\label{table:objective}
\centering
\small{
\begin{tabular}{c|cc|cccc|cccc}
% ---- Top group row: suppress all verticals via \multicolumn alignment ----
\multicolumn{1}{c}{} % no right rule after Method on this row
 & \multicolumn{2}{c}{\textbf{Speech}}
 & \multicolumn{4}{c}{\textbf{Music}}
 & \multicolumn{4}{c}{\textbf{Sound Effect}} \\
\Xhline{3\arrayrulewidth}

% ---- Hz row: draw internal verticals using | in the multicolumns ----
\multirow{2}{*}{\textbf{Method}}
 & \multicolumn{2}{c|}{4kHz \quad 8kHz}
 & \multicolumn{2}{c}{4kHz} & \multicolumn{2}{c|}{8kHz}
 & \multicolumn{2}{c}{4kHz} & \multicolumn{2}{c}{8kHz} \\ \cline{2-11}

% ---- Metric row ----
 & LSD ↓ & LSD ↓
 & LSD ↓ & FD ↓ & LSD ↓ & FD ↓
 & LSD ↓ & FD ↓ & LSD ↓ & FD ↓ \\
\Xhline{3\arrayrulewidth}

% ---- Data rows ----
%model name
% & 0.00 & 0.00
% & 0.00 & 0.00 & 0.00 & 0.00
% & 0.00 & 0.00 & 0.00 & 0.00 \\
Unprocessed
 & 2.89 & 2.49
 & 3.68 & 138.09 & 2.68 & 106.46
 & 3.30 & 110.25 & 2.39 & 64.08 \\

VAE (recon)
 & 0.87 & 0.81
 & 1.13 & 18.92 & 1.06 & 17.30
 & 1.13 & 13.47 & 1.08 & 11.53 \\

\hline

AudioSR \cite{audiosr}
 & 1.46 & 1.26
 & 2.09 & 32.52 & 1.88 & 25.93
 & 1.85 & 39.69 & 1.68 & 28.54 \\

FlashSR \cite{flashsr}
 & 1.47 & 1.15
 & 1.76 & 37.79 & 1.69 & 32.08
 & 1.81 & 41.32 & 1.89 & 36.13 \\

 \hline

SAGA-SR
 & \textbf{1.28} & \textbf{1.07}
 & \underline{1.64} & \textbf{23.87} & \textbf{1.45} & \textbf{20.44}
 & \underline{1.65} & \textbf{26.32} & \textbf{1.43} & \textbf{21.86} \\

w/o text
 & \underline{1.32} & \underline{1.11}
 & \textbf{1.63} & \underline{30.14} & \textbf{1.45} & 25.34
 & \textbf{1.60} & \underline{29.00} & \textbf{1.43} & 23.94 \\

w/o roll-off
 & 1.57 & 1.43
 & 2.16 & 35.99 & 1.78 & \underline{23.5}
 & 2.19 & 33.07 & 1.75 & \underline{22.4} \\

\Xhline{3\arrayrulewidth}

\end{tabular}
}
\end{table*}

Table \ref{table:objective} shows the results of the objective evaluation. SAGA-SR achieves state-of-the-art performance across all metrics and test cases, demonstrating the effectiveness of the proposed method. Compared to its variant without the spectral roll-off embedding, SAGA-SR consistently achieves better performance across all metrics and test cases. This indicates that the spectral roll-off embedding enhances the model’s ability to handle varying cutoff frequencies and guides it to generate outputs with the desired high-frequency energy. When compared with its variant without the text embedding, SAGA-SR achieves superior performance in the speech SR task. In the music and sound effect SR tasks, SAGA-SR and its variant achieve comparable performance in terms of LSD. On the other hand, SAGA-SR consistently outperforms its variant in FD. These results demonstrate that, while the spectral roll-off embedding improves spectral alignment with ground-truth audio, the text embedding is essential for generating outputs that are more plausible and perceptually aligned with the reference audio.

\subsection{Subjective Evaluation}

\begin{figure*}[!t]
\centering
\begin{minipage}{1.0\textwidth}
  \centering
  \centerline{\includegraphics[width=\textwidth]{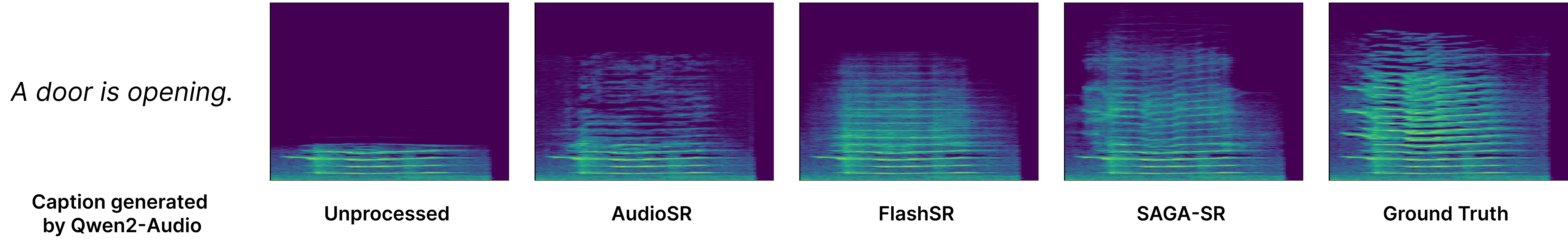}}
\end{minipage}
\caption{Spectrogram of compared models.}
\label{fig:spec}
\end{figure*}

\begin{figure}[!t]
\centering
\begin{minipage}{0.35\textwidth}
  \centering
  \centerline{\includegraphics[width=\textwidth]{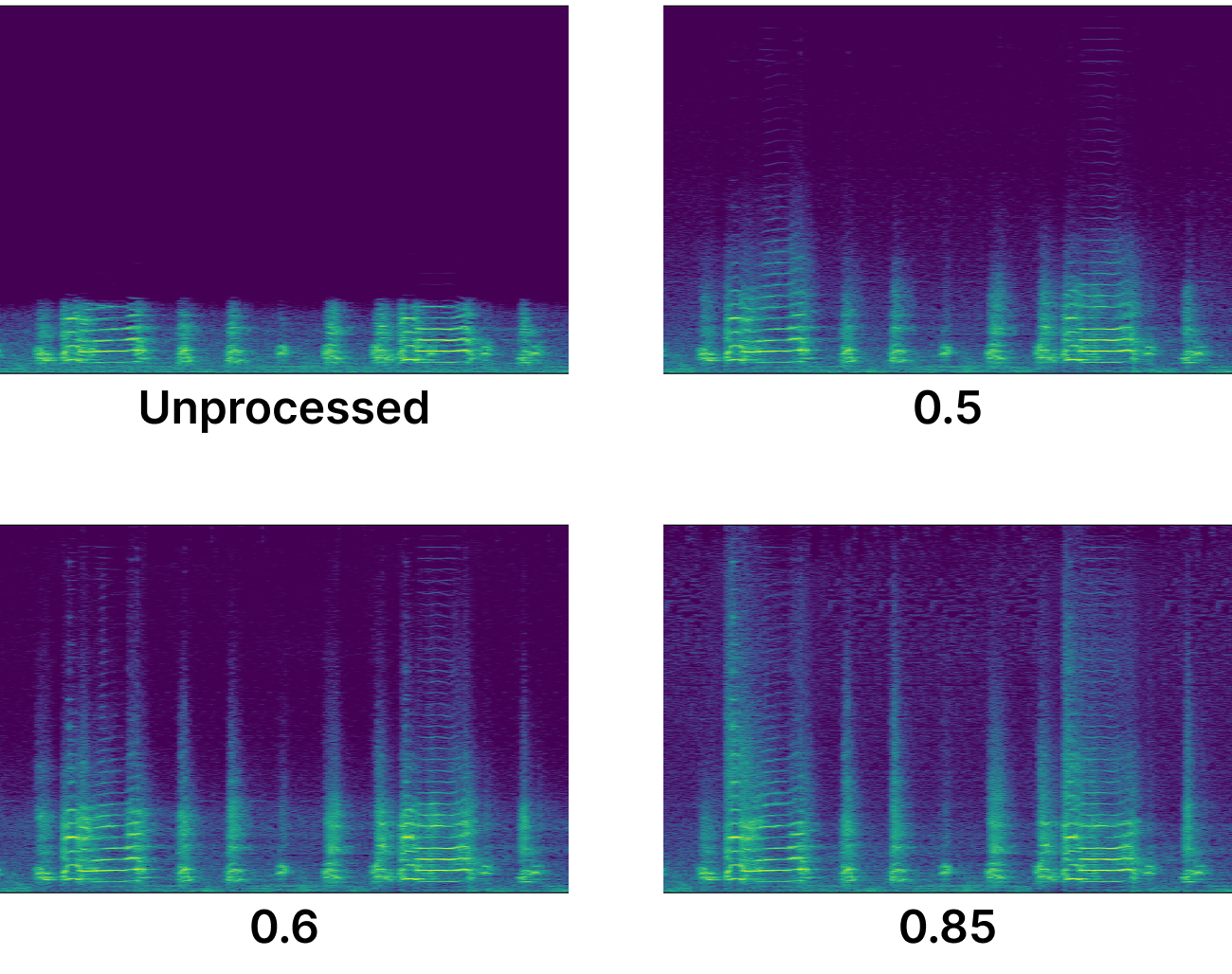}}
\end{minipage}
\caption{Generated samples with varying scales of the target normalized roll-off frequency}
\label{fig:roll}
\end{figure}

Table \ref{table:subjective} presents the results of the subjective evaluation. SAGA-SR achieves the highest scores across all test cases. Figure \ref{fig:spec} shows a comparison of the STFT spectrograms for audio generated by different models. We found that AudioSR and FlashSR exhibit high variance in their outputs and often produce audio lacking sufficient high-frequency content. In contrast, SAGA-SR demonstrates better consistency in reconstructing high-frequency components, due to the explicit guidance provided by the spectral roll-off embedding. Moreover, we observed that AudioSR and FlashSR often generate audio that is not semantically aligned with the low-resolution input audio. Specifically, AudioSR tends to produce outputs with excessive sibilance, while FlashSR often fails to generate detailed harmonic structures, as illustrated in Figure \ref{fig:spec}. By incorporating text embeddings, SAGA-SR is able to generate semantically aligned audio with realistic harmonic detail.

As shown in Figure \ref{fig:roll}, SAGA-SR allows the user to control the high-frequency energy of the generated audio by adjusting the target normalized roll-off frequency, a single scalar condition. This control not only reduces the variance in perceptual quality but also influences acoustic properties such as timbre.

\section{Conclusions}
\label{sec:conclusions}
We propose SAGA-SR, a versatile audio super-resolution model that integrates semantic and acoustic conditioning into a DiT backbone trained with a flow matching objective. Text embeddings improve semantic alignment, while spectral roll-off embeddings enhance robustness and controllability in high-frequency reconstruction. Both objective and subjective evaluations show that SAGA-SR outperforms previous methods across all tasks and metrics.

Despite its effectiveness, SAGA-SR has limitations that warrant future work. First, upsampling audio with multiple overlapping sources remains challenging. Future work could scale data and model capacity or develop improved captioning methods to accurately describe all sound sources. Second, low-frequency replacement post-processing may introduce unnatural connections between high and low frequency components. A potential direction is to develop a VAE conditioned on the low-resolution waveform to achieve smoother integration.

\begin{table}[t]
\caption{Subjective evaluation results.}
\label{table:subjective}
\centering
\small{
\begin{tabular}{c|c|c|c}
\Xhline{3\arrayrulewidth}

% ---- Metric row ----
Method
 & Speech 
 & Music  & Sound Effect \\
\Xhline{3\arrayrulewidth}

% ---- Data rows ----
%model name & 0.00 & 0.00 & 0.00  \\
Unprocessed & 1.81 & 1.66 & 1.77  \\
Ground Truth & 4.23 & 3.93 & 4.18  \\
\hline
AudioSR \cite{audiosr} & 3.26 & 2.94 & 3.03  \\
FlashSR \cite{flashsr} & 3.45 & 3.46 & 3.34  \\
SAGA-SR & \textbf{3.70} & \textbf{3.65} & \textbf{3.88}  \\
\Xhline{3\arrayrulewidth}

\end{tabular}
}
\end{table}

%\bibliographystyle{IEEEbib}
%\bibliographystyle{ieeetr.bst}
\bibliographystyle{IEEEtran}
\bibliography{refs}

\end{document}